# Orientation-dependent hyperfine structure of polar molecules in a rare-gas matrix: a scheme for measuring the electron electric dipole moment

A.C. Vutha
*Department of Physics, University of Toronto*

M. Horbatsch and E.A. Hessels*
*Department of Physics and Astronomy, York University*

Because molecules can have their orientation locked when embedded into a solid rare-gas matrix, their hyperfine structure is strongly perturbed relative to the freely rotating molecule. The addition of an electric field further perturbs the structure, and fields parallel and antiparallel to the molecular orientation result in different shifts of the hyperfine structure. These shifts enable the selective detection of molecules with different orientations relative to the axes of a rare-gas crystal, which will be an important ingredient of an improved electron electric dipole moment measurement using large ensembles of polar molecules trapped in rare-gas matrices.

## I. INTRODUCTION

The electron electric dipole moment ($d_e$) is a model-independent probe of parity and time reversal violation at energies beyond the reach of particle colliders. The typical experimental signature of $d_e$ is the precession of electron spins in the presence of a laboratory electric field. A number of present-day measurements of $d_e$ use heavy diatomic polar molecules, as the interaction between $d_e$ and laboratory electric fields is strongly amplified due to relativistic effects in such molecules. Effectively, the electron spin interacts with an electric field $\mathcal{E}_{\text{eff}}$ that is internal to the molecule, and which can be $\sim 10$ GV/cm or larger in appropriately chosen molecules.

The sensitivity of a statistically-limited measurement of $d_e$ using spin precession is

$$\delta d_e = \frac{\hbar}{2\mathcal{E}_{\text{eff}}\sqrt{N}t_P}. \quad (1)$$

Here, $t_P$ is the precession time during each measurement cycle (limited by the coherence time for the precession), and $N$ is the number of molecules observed to precess multiplied by the number of measurement cycles performed. As $\mathcal{E}_{\text{eff}}$ only differs by an order of unity factor for the heavy molecules used in most experiments measuring $d_e$, it is evident that large $N$ and/or $t_P$ are necessary for significantly improving the precision of searches for $d_e$.

Recently, we have proposed [1] a measurement of $d_e$ using molecules trapped in a rare-gas solid, using a method which we refer to as EDM³ (Electric Dipole Measurements using Molecules in a Matrix). The number of molecules that can be trapped in the solid is large ($\approx 10^{10}$ to $10^{16}$), and, given that the precision of all recent measurements [2–4] is limited by statistics, the larger number could lead to an improvement in measurement accuracy of up to several orders of magnitude. Current measurements of $d_e$ [2, 3] are consistent with zero, and set a 90% confidence interval of $|d_e| < 7\times 10^{-29}$ $e$ cm. The Standard Model predicts that $d_e$ is probably of order $10^{-40}$ $e$ cm [5–7], whereas most extensions of the Standard Model that account for dark matter and the asymmetry between matter and antimatter in the universe (for example, supersymmetric theories [8, 9]) predict a much larger value for $d_e$. A stronger limit (or a nonzero measurement of $d_e$) is necessary to guide Standard-Model extensions.

In addition to allowing for large $N$ and $t_P$, a rare-gas matrix has the advantage, for some molecules in some rare-gas solids (e.g., CO in Ar [10]), that the molecules align themselves along the axes of the rare-gas crystal. Such alignment is achieved without the necessity of an external electric field, and the alignment changes the usual rotational motion of the molecule into a librational motion about one of these axes. These oriented molecules are ideal for measurements of $d_e$, since the measurement would be free from the systematic effects associated with the applied electric field.

In order to use oriented molecules for a measurement of $d_e$, however, it is necessary to experimentally distinguish between molecules with opposite orientations that are contained within the measurement volume. If the (equal and opposite) $d_e$-dependent signals from these molecules are not separately measured, the net signal from the ensemble does not provide any information about $d_e$. In this paper, we point out a new effect – an orientation-dependent shift in the hyperfine structure of the molecules – which can be used to distinguish between molecules with opposite orientations that are trapped within the same matrix.

The purpose of this paper is to describe the ground-state hyperfine structure of these matrix-oriented molecules, and to describe the Stark shifts of the hyperfine structure. This work focuses on ²Σ ground states, and uses the ¹³⁸Ba¹⁹F molecule as an example, since this molecule has been identified as an excellent candidate for measurements of $d_e$ [11, 12], and our modelling [13] shows that it is aligned perpendicularly to the faces of the cubic lattice of solid Ar. This paper also describes how the Stark shifts within the matrix allow for a scheme for measuring $d_e$ using BaF in an Ar crystal.

---

* hessels@yorku.ca

The paper is organized as follows. In Section II, the Stark shift and hyperfine structure for a freely rotating molecule is detailed. The concepts introduced in this section are central to understanding the Stark shift of molecules that are oriented within a rare-gas matrix, which is discussed in Section III, where we describe how the hyperfine structure is affected by the direction of the electric field relative to the orientation of the matrix-trapped molecule. In Section IV, the additional effects of electronically and vibrationally excited states on the hyperfine structure of matrix-isolated molecules in an electric field are described. The usefulness of the shifts described in this work for a measurement of the electron electric dipole moment are detailed in Section V.

## II. HYPERFINE STRUCTURE AND STARK SHIFTS FOR THE FREE MOLECULE

The $^{138}$Ba and $^{19}$F isotopes have nuclear spins of 0 and $\frac{1}{2}$, respectively. The molecule has one unpaired electron spin, and the ground X $^2\Sigma^+$ electronic state therefore has both total nuclear spin $\vec{I}$ and total electron spin $\vec{S}$ of $\frac{1}{2}$. The X $^2\Sigma^+$ state hyperfine structure has been precisely measured [14–17] and is described by the effective Hamiltonian

$$H_{\text{hfs}}^{\text{eff}} = b\vec{I} \cdot \vec{S} + cI_z S_z. \tag{2}$$

Here, the components of $\vec{S}$ and $\vec{I}$ are relative to the internuclear axis of the molecule (with the F-to-Ba direction along $\hat{z}$). The hyperfine constants $b$ and $c$ are derivable from the hyperfine Hamiltonian and the molecular wavefunctions of the ground state, but, in practice, are obtained by fitting to the observed hyperfine structure.

This effective Hamiltonian, along with the effective Hamiltonian for describing rotations:

$$H_{\text{rot}}^{\text{eff}} = BN^2 - DN^4 + \gamma \vec{N} \cdot \vec{S} + \delta N^2 \vec{N} \cdot \vec{S}, \tag{3}$$

describes the energy levels for the $X^2\Sigma^+$ state of the $^{138}$Ba$^{19}$F molecule. Here the total angular momentum of the molecule is $\vec{F}$, with $\vec{J} = \vec{F} - \vec{I}$ being the angular momentum excluding nuclear spin, and $\vec{N} = \vec{J} - \vec{S}$ being the angular momentum excluding both nuclear and electron spin. The matrix elements needed to evaluate these effective Hamiltonians are detailed in Ref. [18], and include matrix elements that mix states of different $j$ and states of different $n$ (where, $j$ and $n$ are the quantum numbers associated with the operators $J^2$ and $N^2$). For the ground electronic and vibrational state of $^{138}$Ba$^{19}$F

(the X $^2\Sigma^+(v=0)$ state), the constants are

$$B = 6\,473.9586(13) \text{ MHz}$$
$$D = 5.5296(13) \text{ kHz}$$
$$\gamma = 80.954(21) \text{ MHz}$$
$$\delta = 0.111(12) \text{ kHz}$$
$$b = 63.509(32) \text{ MHz}$$
$$c = 8.224(58) \text{ MHz}. \tag{4}$$

The ground rotational state ($n$=0) has hyperfine states with total angular momentum of $f$=0 and $f$=1. Both $f$ and its projection $m_f$ are good quantum numbers, and the $m_f$=0 and $m_f$=±1 levels for $f$=1 are degenerate. The separation between $f$=0 and $f$=1 is $b+c/3$=66.25 MHz, as shown in Fig. 1(a).

The X $^2\Sigma^+(v=0)$ ground state of BaF has a permanent electric dipole moment [14] of $\mu_e$=3.170(3) debye. The Stark Hamiltonian in an electric field $\vec{\mathcal{E}} = \mathcal{E}_Z \hat{Z}$ is

$$H_{\text{St}} = -\vec{\mu}_e \cdot \vec{\mathcal{E}} = -\mu_e \mathcal{E}_Z \cos\Theta, \tag{5}$$

where $\Theta$ is the angle between $\vec{\mu}_e$ (which is in the $\hat{z}$ direction) and $\vec{\mathcal{E}}$ (which is in the $\hat{Z}$ direction). (Here the frame fixed to the molecule is denoted by $\hat{x}$, $\hat{y}$, and $\hat{z}$; and the laboratory-fixed frame is denoted by $\hat{X}$, $\hat{Y}$, and $\hat{Z}$.) The matrix elements of $H_{\text{St}}$ for the hyperfine states are

$$\langle njfm_f|H_{\text{St}}|n'j'f'm'_f\rangle = \mu_e \mathcal{E}_Z (-1)^{1-m_f} \xi_{ff'}\xi_{jj'}\xi_{nn'}$$
$$\delta_{m_f m'_f} \begin{pmatrix} f & 1 & f' \\ -m_f & 0 & m_f \end{pmatrix} \begin{pmatrix} n & 1 & n' \\ 0 & 0 & 0 \end{pmatrix} \begin{Bmatrix} f & 1 & f' \\ j' & \frac{1}{2} & j \end{Bmatrix} \begin{Bmatrix} j & 1 & j' \\ n' & \frac{1}{2} & n \end{Bmatrix}, \tag{6}$$

where 3j and 6j symbols are employed, and $\xi_{ss'} = (-1)^{s+s'}\sqrt{(2s+1)(2s'+1)}$. The Stark energy levels are obtained by diagonalizing $H_{\text{hfs}}^{\text{eff}} + H_{\text{rot}}^{\text{eff}} + H_{\text{St}}$. Since $H_{\text{St}}$ can cause substantial mixing of $n$, a large number of $n$ states must be included to ensure convergence. (In this work we include all $n \leq 40$.) The lowest energies from this diagonalization are shown (as solid lines) in Fig. 2. The diagonalization leads to eigenstates for which (because of azimuthal symmetry about the $Z$ direction) $m_f$ is still a good quantum number; however, $f$ is no longer a good quantum number. For ease of notation, despite the mixing of $f$ eigenstates caused by Eq. (6), we continue to label hyperfine states with the $f$ value of the state to which they are adiabatically connected in the field-free limit. Note that with the presence of the electric field, $m_f$=0 is no longer degenerate with $m_f$=±1, but the $m_f$=−1 and $m_f$=+1 states remain degenerate.

For small electric fields, the Stark shift shows the quadratic dependence on electric field that is expected from second-order perturbation theory:

$$\Delta_{\text{St}}^{\text{small}} \approx \frac{\mu_e^2 \mathcal{E}_Z^2}{6B}. \tag{7}$$

This small-field approximation (dashed orange line) is compared to the full diagonalization (solid blue line) in Fig. 2(b), where the $f$=0 and $f$=1 hyperfine structure is also resolved.

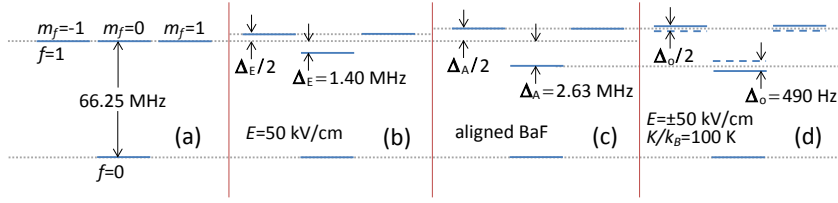

FIG. 1. The hyperfine structure for the lowest-energy states of $^{138}$Ba$^{19}$F (relative to the lowest hyperfine state; not to scale). Panel (a) shows the hyperfine states for the free molecule. The positions of the $f=1$ states (relative to the $f=0$ state) is shifted by an electric field, (b). The hyperfine structure for a perfectly aligned molecule is shown in (c). The combination of a Devonshire potential and an electric field leads to slightly different Stark shifts, (d), for molecules oriented parallel (dashed lines) and anti-parallel (solid lines) to the electric field.

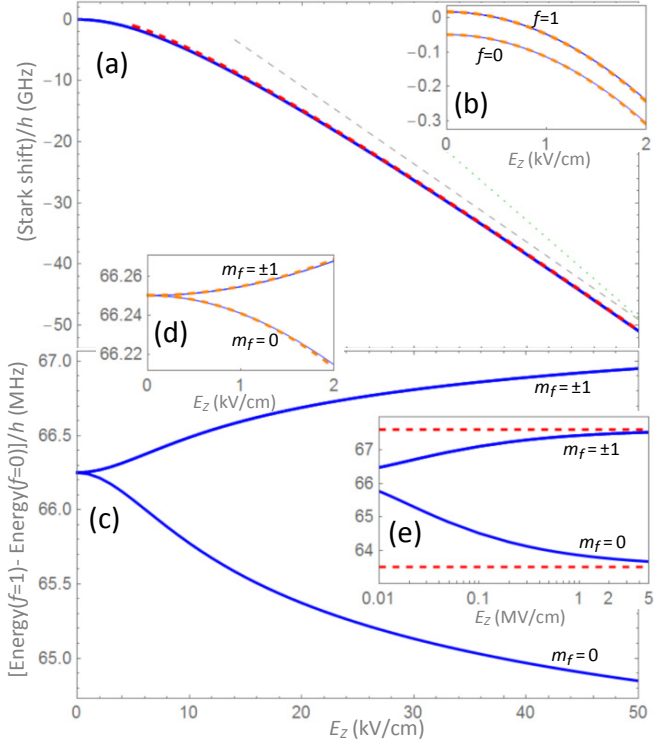

FIG. 2. The Stark shift for the lowest-energy states of $^{138}$Ba$^{19}$F. In (a), the shift (from the centroid of the $n=0$ zero-electric-field energies) obtained from a full diagonalization (solid blue line) is compared to the large-field approximation of Eq. (8) (thick dashed red line). Also shown in (a) are the linear Stark-shift rate at 50 kV/cm (thin gray dashed line) and the linear Stark-shift rate for very large fields of $-\mu_e$ (dotted green line). A magnification of the low-field part of this plot is shown in (b), which shows that the quadratic approximation of Eq. (7) (dashed orange lines) works well for small fields. The difference between the Stark shifts for $f=1$ and $f=0$ are shown with a much larger magnification in (c). A magnification of the low-field part of this plot, (d), shows that the quadratic approximation (dashed orange line) of Eqs. (9), (10) and (11) works well at low fields. At even larger fields, (e), the separation between the $f=1$ $m_f=0$ and $m_f=\pm 1$ states approaches the separation of $c/2$ expected for perfectly aligned molecules (dashed lines).

For larger fields, $\vec{\mathcal{E}}$ significantly orients $\vec{\mu}_e$. A fully oriented dipole would have a Stark-shift rate of $-\mu_e$, and this rate is shown as a dotted line in Fig. 2(a). Even at a field of 50 kV/cm, the shift rate is only 80 % of this rate (as shown by the thin dashed line), indicating that the component of $\vec{\mu}_e$ along $\hat{Z}$ (on average) is 0.80 $\mu_e$. At this field, the molecule no longer rotates, but is instead in a librational state, in which the direction of $\vec{\mu}_e$ performs angular oscillations about $\hat{Z}$. The wavefunction of the ground librational state determines the degree of orientation, with larger fields restricting the oscillation to orientations closer to $\hat{Z}$. An analysis of the librational states [19] indicates that at larger fields the Stark shifts can be approximated by

$$\Delta_{\text{St}}^{\text{large}} = -\mu_e \mathcal{E}_Z + \sqrt{2B\mu_e \mathcal{E}_Z} - B/2, \qquad (8)$$

as shown by the dashed red line in Fig. 2(a).

Equations (7) and (8) are only approximate. In particular, the equations do not include the weak $f$ and $m_f$ dependence of the Stark shifts. This dependence is shown in Fig 2(c). The relative positions of the $f, m_f$ states for 50 kV/cm is also shown in Fig. 1(b). Since the Stark matrix elements of Eq. (6) depend on $m_f$, the Stark shifts also depend on $m_f$. In the quadratic-shift regime, this $m_f$ dependence reduces to the usual combination of a scalar and a tensor shift:

$$\Delta_{\text{St}}^{\text{small}} = -\frac{\mathcal{E}_Z^2}{2}\left[\alpha^{(0)} + \alpha^{(2)} \frac{3m_f^2 - f(f+1)}{f(2f-1)}\right]. \qquad (9)$$

Additionally, since the hyperfine energies and the matrix elements of $H_{\text{hfs}}^{\text{eff}}$ depend on $f$, the exact Stark shift also depends (weakly) on $f$. The scalar and tensor Stark shift rates are complicated by the fact that the off-diagonal matrix elements of $H_{\text{hfs}}^{\text{eff}}$ are of similar size to the matrix elements of $H_{\text{St}}$. As a result, higher than second order in perturbation theory (similar to [20]) is needed to obtain the quadratic shift rates. We determine the $f$ and $m_f$ dependence of our shift rates directly from the energies obtained from our diagonalization:

$$\Delta_{\text{St}(f=1,m_f=\pm 1)}^{\text{small}} - \Delta_{\text{St}(f=0)}^{\text{small}} = 4.66 \text{ kHz/(kV/cm)}^2 \mathcal{E}_Z^2$$

$$\Delta_{\text{St}(f=1,m_f=\pm 1)}^{\text{small}} - \Delta_{\text{St}(f=1,m_f=0)}^{\text{small}} = 14.01 \text{ kHz/(kV/cm)}^2 \mathcal{E}_Z^2$$

$$\Delta_{\text{St}(f=1,m_f=0)}^{\text{small}} - \Delta_{\text{St}(f=0)}^{\text{small}} = -9.35 \text{ kHz/(kV/cm)}^2 \mathcal{E}_Z^2. \quad (10)$$

These quadratic shift rates are shown as dashed orange lines in Fig. 2(d), and are due to scalar and tensor shift rates of

$$\alpha^{(0)}(f=1) - \alpha^{(0)}(f=0) = 0.02 \text{ kHz/(kV/cm)}^2$$
$$\alpha^{(2)}(f=1) = -9.34 \text{ kHz/(kV/cm)}^2. \quad (11)$$

At the larger fields shown in Fig. 2(c), the quadratic approximation no longer holds. In the limit of very large fields, the molecule becomes almost perfectly aligned, and the $f=1, m_f=0$ to $f=1, m_f=\pm1$ energy separation approaches the perfectly aligned value of $c/2$ as indicated by the dashed red lines in Fig. 2(e). The hyperfine structure for perfect alignment is also shown in Fig. 1(c).

## III. STARK SHIFT FOR MOLECULES ORIENTED BY A RARE-GAS MATRIX

In the previous section, the BaF molecules were oriented by large laboratory fields. Another way to orient the molecules is to isolate them in a rare-gas crystal. The BaF molecule strongly prefers the six orientations normal to the cubic structure of the crystal, which we here assume to be along the $\pm\hat{X}$, $\pm\hat{Y}$, and $\pm\hat{Z}$ directions. We note that the true eigenstates of the system are linear combinations of these six orientations, due to tunneling between the angular potential minima. However, the tunneling matrix elements are extremely small for the case of deep potentials [21, 22]. The potential that constrains the molecular orientation in the matrix has octahedral symmetry, and, to lowest order, is given by a Devonshire potential [23, 24]

$$H_{\text{Dev}} = -\frac{K}{8}\left(3 - 30\cos^2\Theta + 35\cos^4\Theta + 5\sin^4\Theta\cos 4\Phi\right)$$
$$= -\frac{K}{3}\left[\sqrt{\frac{10\pi}{7}}\left(Y_{4,-4} + Y_{4,4}\right) + \sqrt{4\pi}Y_{4,0}\right]. \quad (12)$$

The constant $K$ determines the strength of the potential and the degree to which it orients the molecule. Our modeling of BaF in an argon matrix [13] indicates that $K/k_B$ is of the order of 100 kelvin, where $k_B$ is the Boltzmann constant. The matrix elements of $H_{\text{Dev}}$ are:

$$\langle njfm_f|H_{\text{Dev}}|n'j'f'm'_f\rangle = K\xi_{ff'}\xi_{jj'}\xi_{nn'}(-)^{-m_f}$$
$$\left[\sqrt{\frac{5}{14}}\begin{pmatrix}f & 4 & f'\\-m_f & 4 & m'_f\end{pmatrix} + \sqrt{\frac{5}{14}}\begin{pmatrix}f & 4 & f'\\-m_f & -4 & m'_f\end{pmatrix}\right.$$
$$\left.+\begin{pmatrix}f & 4 & f'\\-m_f & 0 & m'_f\end{pmatrix}\right]\begin{pmatrix}n & 4 & n'\\0 & 0 & 0\end{pmatrix}\begin{Bmatrix}f & 4 & f'\\j' & \frac{1}{2} & j\end{Bmatrix}\begin{Bmatrix}j & 4 & j'\\n' & \frac{1}{2} & n\end{Bmatrix}, \quad (13)$$

where the $\xi$ factors were defined in Section II.

The Stark shifts within the Devonshire potential can be obtained by diagonalizing $H_{\text{hfs}}^{\text{eff}} + H_{\text{rot}}^{\text{eff}} + H_{\text{St}} + H_{\text{Dev}}$. As in the previous section, $f$ is no longer a good quantum number. In addition, since $H_{\text{Dev}}$ does not have azimuthal symmetry, $m_f$ is also not an exact quantum number. As was done in the previous section, we continue to label hyperfine states with the $f$ and $m_f$ values of the state to which they are adiabatically connected in the field-free limit. Also as in the previous section, the $|f=1, m_f=\pm1\rangle$ states are no longer degenerate with the $|f=1, m_f=0\rangle$ state; however, the $|f=1, m_f=\pm1\rangle$ states remain degenerate.

Fig. 3 gives the results of such a diagonalization of $H_{\text{hfs}}^{\text{eff}} + H_{\text{rot}}^{\text{eff}} + H_{\text{St}} + H_{\text{Dev}}$ for a Devonshire potential with $K/k_B$=100 kelvin, and with an electric field along the $\hat{Z}$ axis which ranges from 1 to 50 kV/cm. Panel (a) of the figure shows the basic structure of the Stark shifts. The molecules locked into the $\pm\hat{Z}$ directions by the matrix experience a Stark shift of about $\mp\mu_e\mathcal{E}_Z$. Close inspection of the figure indicates that the shift is slightly smaller than $\mp\mu_e\mathcal{E}_Z$, which is a result of the ground-state librational motion for the molecules. The molecules oriented in the plane perpendicular to $\vec{\mathcal{E}}$ show negligible shifts.

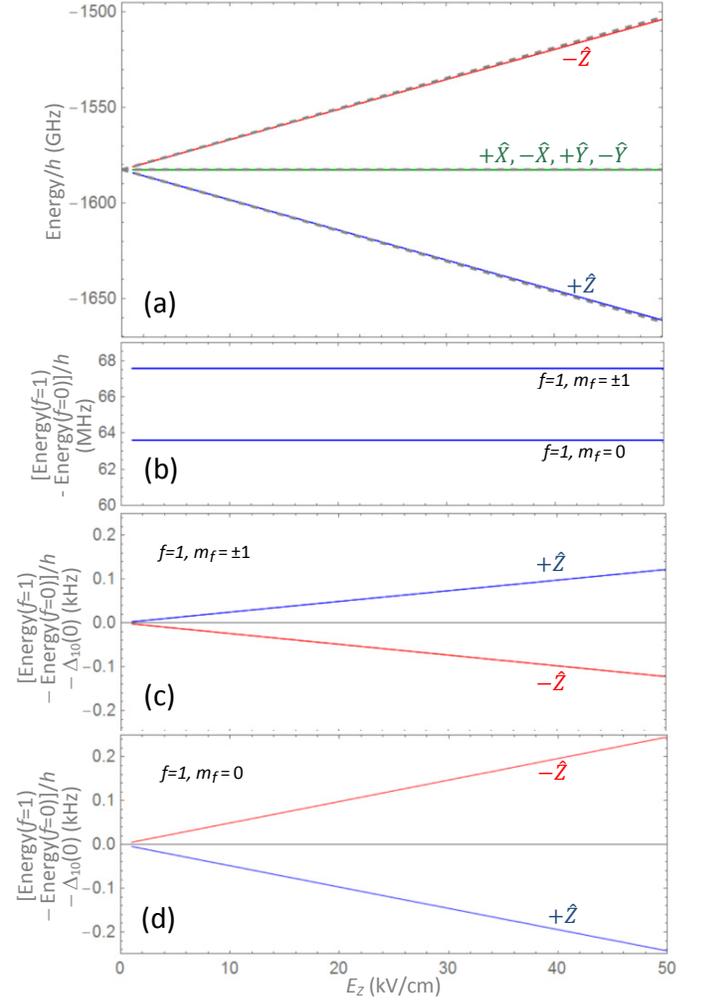

FIG. 3. The Stark shifts for the lowest-energy states of $^{138}$Ba$^{19}$F within a matrix (which is modeled using a Devonshire potential with $K/k_B =100$ K) are shown. Panel (a) shows that the $\pm\hat{Z}$ oriented molecules exhibit Stark shifts of nearly $\mp\mu_e\mathcal{E}_Z$ (dashed gray lines), whereas the $\pm\hat{X}$ and $\pm\hat{Y}$ oriented molecules have an almost zero Stark shift. On a scale with a larger magnification in (b), where energy differences between the $f=1$ and $f=0$ states are shown, the hyperfine structure is evident. Unlike the free molecule of Fig. 2(c), the hyperfine structure is approximately independent of electric field. However, using a magnified scale, (c) and (d), a linear shift of the hyperfine structure is evident. For these graphs, the average of the $+\hat{Z}$ and $-\hat{Z}$ values for the intervals ($\Delta_{10}(0)$) are subtracted. This small linear shift is has the opposite sign for $+\hat{Z}$ and $-\hat{Z}$ orientations of the molecule, and the different shifts can be used to selectively address one orientation of the molecule with rf fields.

The hyperfine structure is not resolved in Fig. 3(a), but, on a magnified scale in Fig. 3(b), the separation of the $f$=1 and $f$=0 states is evident. Unlike in the case for the free molecule (Fig. 2(b)), the hyperfine structure of Fig. 3(b) is nearly independent of electric field. The near independence is a result of the fact that the molecule is almost perfectly aligned by the Devonshire potential, and thus its alignment is almost independent of the applied electric field.

Fig. 3(b) shows the hyperfine structure for all six orientations of the molecule, and does not resolve any differences between the orientations. On the largely magnified scale of Fig. 3(c) and (d), however, a small dependence of the hyperfine structure on electric field becomes evident. The hyperfine intervals are changing linearly with electric field, and show the opposite sign shifts for molecules aligned along the $+\hat{Z}$ and $-\hat{Z}$ directions. For the $+\hat{Z}$ orientation, the electric field works to strengthen the molecular alignment (increase the strength of the potential well and therefore reduce that range of the librational motion) and therefore the hyperfine structure becomes slightly closer to the perfectly aligned structure of Fig. 1(c). For the $-\hat{Z}$ orientation, the opposite occurs, with the confining potential being reduced by the electric field and the hyperfine separations therefore become slightly further from the perfectly aligned structure. This effect of an electric field on the hyperfine structure of a matrix-oriented BaF molecule is also shown schematically in Fig. 1(d).

The linear variation of the hyperfine energy difference with electric field for molecules oriented by a matrix is the main result of this work. For the particular example of BaF in a Devonshire potential with $K/k_B$=100 kelvin and an applied electric field of 50 kV/cm, the energy difference is 490 Hz, as shown at the right of Fig. 3(d) (and in Fig. 1(d)). As a result, oppositely oriented molecules have different hyperfine splittings. This difference allows for driving hyperfine transitions that are orientation specific in the presence of an applied electric field. This energy difference increases linearly with $\mathcal{E}_Z$, and scales approximately as $\mathcal{E}_Z\mu_e c\sqrt{B/K^3}$ (where $c$ and $B$ are the constants from Eq. (4)).

In addition to looking at the energies from the diagonalization of $H_{\text{hfs}}^{\text{eff}}+H_{\text{rot}}^{\text{eff}}+H_{\text{St}}+H_{\text{Dev}}$, it is also instructive to look at the decomposition of the eigenstates. As indicated above, $m_f$ is not formally a good quantum number for a molecule that experiences both a Devonshire potential and an applied electric field. However, inspection of the eigenvectors indicates that there is no significant admixture of states with other $m_f$ values. That is, $m_f$ is very nearly a good quantum number. To understand this result, we observe that the Devonshire potential of Eq. (12) becomes nearly independent of $\Phi$ for the small $\Theta$ that the molecule is restricted to for the ground librational state. Additionally, since Eq. (13) only allows admixtures of other $m_f$ states with $\Delta m_f$=$\pm 4$, only states at much higher energy (highly-excited librational states for which the orientation of the of the molecule is less strongly restricted) can lead to $\Delta m_f\neq 0$ admixtures.

## IV. THE EFFECT OF ELECTRONICALLY AND VIBRATIONALLY EXCITED STATES

A second mechanism can lead to shifts of the hyperfine structure that are also linear with $\mathcal{E}_Z$, and therefore have the same character as the shifts shown in Fig. 3(c) and (d). These shifts come from the polarization of the ground state X $^2\Sigma^+(v$=0$)$ due to the applied electric field. For a molecule that is aligned along either the $+\hat{Z}$ or $-\hat{Z}$ direction by the matrix, the ground state $|g\rangle$ is perturbed by the field:

$$|g'\rangle = |g\rangle \pm \sum_{\text{e}} \frac{\langle \text{e}|ez\mathcal{E}_Z|\text{g}\rangle}{E_{\text{g}}-E_{\text{e}}}|\text{e}\rangle, \qquad (14)$$

where the sum is over all vibrationally and electronically excited states $|\text{e}\rangle$. The polarization causes the hyperfine matrix elements that determine the hyperfine structure to change from $\langle\text{g}|H_{\text{hfs}}|\text{g}\rangle$ to $\langle\text{g}'|H_{\text{hfs}}|\text{g}'\rangle$. This leads to additional energy shifts (due to polarization) of the form

$$\Delta E = \pm 2\sum_{\text{e}} \frac{\langle\text{g}|H_{\text{hfs}}|\text{e}\rangle\langle\text{e}|z|\text{g}\rangle}{E_{\text{g}}-E_{\text{e}}}e\mathcal{E}_Z, \qquad (15)$$

where the $\pm$ sign depends on the direction of the molecular axis relative to the direction of the field. Neither of the matrix elements in Eq. (15) can be calculated without the full wavefunctions for the ground and excited molecular states. For electronically excited states $|\text{e}\rangle$, one would expect that the hyperfine matrix element would be of the same order as the $\langle g|H_{\text{hfs}}|g\rangle$ matrix elements that determine the X $^2\Sigma^+(v=0)$ hyperfine structure. Since the matrix element of $z$ and the energy differences are of order of an atomic unit, it could be expected that electronic states would contribute of order (10 MHz)$(\mathcal{E}_Z/\mathcal{E}_{\text{au}})$, where $\mathcal{E}_{\text{au}}=5$ GV/cm is the atomic unit for electric field. This leads to a linear shift of the hyperfine structure that is on the order of 100 Hz for a 50 kV/cm electric field, only slightly smaller than the shifts indicated by the effect of the previous section.

For vibrationally excited states, X $^2\Sigma^+(v$>0$)$, the energy denominator in Eq. (15) is about an order of magnitude smaller than in the case of electronically excited states, and the $z$ matrix elements are about two orders of magnitude smaller, making the expected contributions from excited vibrational states an order of magnitude smaller than those from excited electronic states.

## V. APPLICATION TO A MEASUREMENT OF THE ELECTRON ELECTRIC DIPOLE MOMENT

The EDM³ method exploits the effect discussed in this work. Here we discuss the EDM³ measurement sequence, which is illustrated in Fig 4. The seven steps of one measurement cycle are illustrated in panels (a) through (g). A small bias magnetic field $B_Z \hat{Z}$ (of about 1 mgauss) is applied during all steps along one of the axes of the argon crystal. The applied electric field is zero throughout the experiment, except for step (f), where the Stark shift of the hyperfine structure is used to distinguish oppositely oriented molecules. We describe the experiment sequence using the specific example of BaF molecules in an Ar matrix. However, we emphasize that the method is applicable to other polar molecules and matrices.

Before the steps shown in Fig 4, a single crystal of argon is made with $^{138}$Ba$^{19}$F molecules embedded in a ratio of, for example, 1(BaF):$10^9$(Ar). At this concentration, there would be $n_{\text{used}} \approx 10^{13}$ BaF molecules per cm$^3$ that are aligned along a single direction. At a temperature of 3 kelvin, all of the atoms will thermalize into the ground electronic state, the ground vibrational state, the ground librational state, as well as the ground state for the molecule's center-of-mass motion.

The four hyperfine states begin with equal populations. In step (a) of the figure, the population is optically pumped to the X $^2\Sigma^+ m_f$=+1 state using $\sigma^+$ circularly-polarized laser light tuned to the A $^2\Pi_{1/2}$ state. Since electronic transitions within the matrix are usually broadened beyond their free-molecule natural width, the hyperfine structure cannot be resolved. However, for $\sigma^+$ polarization, the X $^2\Sigma^+ m_f$=+1 state is a dark state (since there are no $\Delta m_f$=+1 transitions available for the $\sigma^+$ laser driving to the A $^2\Pi_{1/2}$ state), and population will accumulate into this state. The $\sigma^+$ laser light does not strongly couple the two $m_f$=0 hyperfine states to the A $^2\Pi_{1/2}$. Therefore, $\sigma^-$ rf fields are used to couple these two $m_f$=0 states to the $m_f$=−1 state. At the end of step (a), almost all of the population is transferred into the X $^2\Sigma^+ m_f$=+1 state.

In step (b) of the figure, this population is transferred into one of the X $^2\Sigma^+ m_f$=0 states using a $\sigma^+$ rf $\pi$ pulse. From this $m_f$=0 state, the population is transferred to the even superposition of the X $^2\Sigma^+ m_f$=−1 and +1 states:

$$|\psi^+\rangle = \frac{|m_f = -1\rangle + |m_f = +1\rangle}{\sqrt{2}} \quad (16)$$

using $\hat{X}$-polarized rf fields, as shown in panel (c). Alternatively, $\hat{Y}$-polarized rf fields would populate the

$$|\psi^-\rangle = \frac{|m_f = -1\rangle - |m_f = +1\rangle}{\sqrt{2}} \quad (17)$$

state.

Beginning after the start rf pulse shown in panel (c) of the figure, then continuing for a time $t_P$, until the stop rf pulse of panel (e), the quantum mechanical phases progress, as shown in (d). The $m_f$=+1 and −1 states have their energies shifted by $+\hbar\omega_P$ and $-\hbar\omega_P$, respectively, due to the applied magnetic field and the effective electric field inside the aligned molecules. For molecules oriented in the $\pm\hat{Z}$ directions, $\omega_P = (g\mu_B B_Z \pm d_e \mathcal{E}_{\text{eff}})/\hbar$ is the precession frequency caused by the interaction of the magnetic moment $g\mu_B$ with the magnetic field, and $d_e$ with the effective electric field. Thus, by the end of the phase evolution step, the state is given by

$$|\psi(t_P)\rangle = \frac{|m_f = -1\rangle e^{i\omega_P t_P} + |m_f = +1\rangle e^{-i\omega_P t_P}}{\sqrt{2}}$$
$$= \cos(\omega_P t_P)|\psi^+\rangle + i\sin(\omega_P t_P)|\psi^-\rangle. \quad (18)$$

The limit on the maximum value of $t_P$ that can be used is the coherence time of the phase precession. This coherence time is limited by the fact that different molecules will see different magnetic fields. Because superconductors can be used to shield magnetic fields at the 3-kelvin temperature used here, and because the argon sample is small, the applied magnetic field can be made to be very uniform over the sample. However, the magnetic field due to neighboring BaF molecules will lead to small variations in magnetic field, which will serve to decohere the phases. This effect is referred to as dipolar relaxation [25], and limits $t_P \lesssim 100\text{ms} \times \frac{n_{\text{used}}}{10^{13}/\text{cm}^3}$.

The stop pulse in (d) uses $\hat{X}$ polarized rf to transfer the $\psi^+$ component of $\psi(t_P)$ to one of the X $^2\Sigma^+ m_f$=0 states, and $\hat{Y}$ polarized rf to transfer the $\psi^-$ component to the other $m_f$=0 state. After this step, the populations in the two $m_f$=0 states, are $s=\sin^2(\omega_P t_P)$ and $c=\cos^2(\omega_P t_P)$. Thus, the two populations now can be used to determine the sine and cosine of phase $\omega_P t_P$, and, therefore, the phase progression can be measured using these populations.

Step (f) of Fig. 4 moves the population from the two X $^2\Sigma^+ m_f$=0 states into the X $^2\Sigma^+ m_f$=±1 states. For this step an electric field is applied, so that the $\pm\hat{Z}$-oriented molecules will have slightly different hyperfine structure splittings (as described in Sections III and IV), allowing for the rf transitions to separately address the $+\hat{Z}$ and $-\hat{Z}$ oriented molecules. A total of four different rf fields are used in this step, all at different frequencies, and with the polarizations indicated in panel (f). For the $-\hat{Z}$ molecules (for which the energy levels and rf transitions are represented by dashed lines and arrows), the population from one of the $m_f$=0 states (the one labeled c in panel (e) is transferred to the $m_f$=+1 (as represented by the orange circle containing a c), and the other $m_f$=0 state (labeled by s) is transferred to the $m_f$=−1 state. The $+\hat{Z}$ molecules (represented by solid lines, solid arrows and purple circles in the figure) make the opposite transitions, with the s population going to $m_f$=+1 and the c population going to $m_f$=−1.

The X $^2\Sigma^+ m_f$=−1 state population is detected in (g), where an applied electric field is no longer needed, by driving the X $^2\Sigma^+(m_f=-1)(v=0)$-to-A $^2\Pi_{3/2}(m_f=-2)(v=0)$ transition with $\sigma^-$ circularly polarized laser light, and observing the resulting fluorescence. This transition is nearly a cycling transition, with, due to a favorable Franck-Condon factor, $\sim 95\%$ of the population returning to X $^2\Sigma^+(v=0)$ state. All of this $\sim 95\%$ of the population can only decay to $m_f=-1$, and the remaining $\sim 5\%$ of the population decays to higher-$v$ states. Thus, even without repumping molecules out of these higher-$v$ states, an average of about 20 photons are emitted from each molecule. With a fluorescence detection efficiency of $\geq 5\%$, the molecules can be detected with near unit efficiency. Following the detection of the X $^2\Sigma^+ m_f=-1$ population, the X $^2\Sigma^+ m_f=+1$ population can be detected in a similar fashion using $\sigma^+$ laser light.

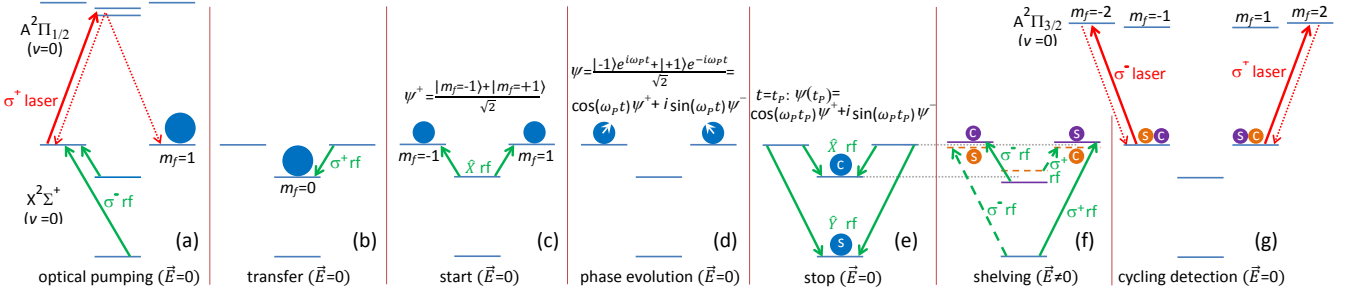

FIG. 4. A scheme for measuring the electron electric dipole moment using $^{138}$Ba$^{19}$F molecules in a rare-gas matrix. The BaF molecules thermalize into the four X $^2\Sigma^+$ hyperfine states shown, and (a) are optically-pumped (along with rf couplings) into the X $^2\Sigma^+ m_f=+1$ state. This population is (b) transferred to an $m_f=0$ state and then (c) into $\psi^+$, the even superposition of the $m_f=-1$ and $m_f=+1$ states. The phases of the two components of $\psi^+$ evolve (d) for a time $t_P$, before the resulting $\psi^+$ and $\psi^-$ components are transferred (e) into the two $m_f=0$ states. From these two states, the population is moved into the $m_f=-1$ and $m_f=+1$ states by appropriate rf fields. This step is performed with an applied electric field $\mathcal{E}_Z \hat{Z}$, so that distinct frequency-resolved transitions can be driven for molecules oriented in the $+\hat{Z}$ and $-\hat{Z}$ directions. These distinct transitions make the scheme insensitive to magnetic fields, while doubling the sensitivity to $d_e$. Finally, (g) the $m_f=-1$ and $m_f=+1$ populations are measured by separately exciting them with $\sigma^-$ and $\sigma^+$ laser transitions, and detecting the resulting fluorescence. The ratio of the fluorescence for the $\sigma^-$ and $\sigma^+$ cases gives a measurement of $d_e$, as described in the text. Also described in the text are nine reversals possible in this scheme that will allow for the measurement and cancellation of systematic effects.

At the end of the measurement cycle, the majority of the molecules will be in X $^2\Sigma^+(v>0)$ states, which have vibrational relaxation times of a few hundred milliseconds. Thus, a delay time of one second between measurement cycles is required to allow the population to relax back to the $v=0$ state. Alternatively, repump lasers could be used to move the populations more quickly to $v=0$. Similarly, the optical-pumping step in (a) will lead to some population in higher-$v$ states. This higher-$v$ population could also be removed with repump lasers, or simply be allowed to relax back to $v=0$ by radiative decay.

The magnetic field $B_Z$ and the precession time $t_P$ can be chosen so that the total precession angle $\omega_P t_P$ is approximately equal to an integer multiple of $2\pi$ plus $\pi/4$. That is,

$$\omega_P t_p \bmod 2\pi = \frac{\pi}{4} + \delta, \qquad (19)$$

where $\delta$ is small, so that $c = \frac{1}{2} - \delta + \mathcal{O}(\delta^2)$ and $s = \frac{1}{2} + \delta + \mathcal{O}(\delta^2)$, with the $\mathcal{O}(\delta^2)$ terms being negligible. Here, $\delta = \delta_B \pm \delta_E$, where $\delta_B$ results from an imperfect magnetic field for the precession time used, and $\delta_E = d_e \mathcal{E}_\text{eff} t_P/\hbar$ results from the electric dipole moment of the electron, with the $\pm$ referring to the molecules oriented in the $\pm\hat{Z}$ directions. The ratio of the population in $m_f=-1$ to that in $m_f=+1$ in figure Fig. 4(f) is then given by $f^+(1-4\delta^+) + f^-(1+4\delta^-)$, where $f^\pm$ is the fraction of molecules oriented in the $\pm\hat{Z}$ direction. In the approximation that there are equal populations in each of these orientations, the contribution from $\delta_B$ cancels in the ratio. This subtraction isolates just the effect of $d_e$, and makes the measurement insensitive to, e.g., $B$-field drifts. Here, the oppositely oriented molecules act as co-magnetometers for each other. Note that this method allows even $^2\Sigma$ molecules, which do not have the $\Omega$-doublet co-magnetometer structure [2, 3, 26], to be effective for high-precision electron electric dipole moment experiments. With the cancellation of $\delta_B$, the ratio becomes $1 - 4d_e\mathcal{E}_Z t_P/\hbar$, and therefore the ratio of the fluorescence seen after the $\sigma^+$ and $\sigma^-$ excitations of Fig. 4(f) gives a measurement of $d_e$.

In case the measurement is not performed perfectly (e.g., different powers for the $\sigma^-$ and $\sigma^+$ lasers in (g), a difference between $f^+$ and $f^-$, or imperfect polarizations in the steps (a) through (f)), many reversals can be made to the scheme depicted in Fig. 4 to determine the imperfections and to cancel their effect on the measurement. These reversals are enumerated below. (1) The direction of the applied magnetic field $B_Z$ can be reversed, by changing the sign of the current that produces the field. (2) The magnetic field can also be reversed by physically rotating the coils that produce the field. (3) The directions of the circular polarizations in (a) can be reversed, to optically pump the population into the X $^2\Sigma^+ m_f=-1$ state. (4) The frequency of the rf in (b) can be tuned to transfer the population to the other $m_f=0$ state. (5) The rf polarization in (c) can be changed to $\hat{Y}$ to excite the atoms into the odd superposition $\psi^-$. (6) The polarizations of the two rf fields in (e) can be changed (from $\hat{X}$ and $\hat{Y}$ to $\hat{Y}$ and $\hat{X}$, respectively) to exchange which of the $m_f=0$ is connected to $\psi^+$ and which is connected to $\psi^-$. (7) The polarizations can be reversed in (f). (8) The direction of the applied electric field can also be reversed in (f). (9) The order in which the $\sigma^+$ and $\sigma^-$ lasers are applied in (g) can also be reversed.

Further systematic tests can be performed by varying (by factors of 10 to 1000) the magnitude of $B_Z$, the magnitude of $\mathcal{E}_Z$, the time taken for each step, the delay between steps, and the BaF:Ar ratio. Possible additional tests of systematics would include embedding more than one type of molecule (with different sensitivities to $d_e$), physically reversing the solid argon sample, or repeating the experiment along the other two axes of the crystal (the $\hat{X}$ and $\hat{Y}$ axes).

The statistical uncertainty $\delta d_e$ of a measurement using the EDM$^3$ method can be obtained from Eq. (1), with $N$ being proportional to the volume $V$ of the argon crystal, to the density of BaF used ($n_{\text{used}}$) and to the number of experiment cycles. The precession time $t_P$ of each measurement cycle is limited by the coherence time, which scales as $1/n_{\text{used}}$. The number of experimental cycles that can be performed within a period $T_{\text{measure}}$ of data collection is $T_{\text{measure}}/t_P$ times the duty cycle $D_C$. Putting these together gives an estimate for $\delta d_e$ that is independent of $n_{\text{used}}$:

$$\delta d_e \approx 1 \times 10^{-34} e\,\text{cm} \sqrt{\frac{1\text{ cm}^3}{V}} \sqrt{\frac{1\text{ month}}{T_{\text{measure}}}}, \qquad (20)$$

where we have assumed $D_C=10\%$, and $n_{\text{used}}$ is adjusted to obtain $t_P=100$ ms.

Of course, systematic uncertainties could be larger than this statistical uncertainty; however, previous measurements have been able to control systematic uncertainties down to the level of statistical uncertainties [2–4]. The EDM$^3$ method has some advantages for further controlling systematic uncertainties. First, the small experimental volume makes it possible to better shield external fields and to have better uniformity of fields. Second, no applied electric field is present during the precession step, and therefore no electric-field systematics (e.g., leakage currents) can be present. Third, the molecules remain stationary during the whole measurement, and therefore there will be no systematics due to the motion of the molecules, such as motional electric or magnetic fields, geometric phases, or nonadiabatic effects from the motion of molecules through laser beams. Fourth, oppositely oriented molecules offer an ideal co-magnetometer for rejecting systematics and isolating the true $d_e$-dependent signal. Fifth, the measurement is performed in a cryogenic environment, which allows for shielding of magnetic fields using superconductors and suppresses thermal voltages and currents. Sixth, the large number of reversals allows for a direct measurement and canceling of many types of systematic effects. Seventh, the large dynamic range on several experimental parameters (e.g., $B_Z$, $t_P$, delay times) allows for further exploration of possible systematic effects. Eighth, the measurement can easily be repeated with a newly-grown crystal with different levels of impurities, imperfections, or with different rare gases or substrates. Ninth, a control molecule (e.g., CaF or YbF) with a different sensitivity to $d_e$ could be used.

## VI. CONCLUSION

We have examined the hyperfine structure of molecules oriented in rare-gas matrices, and found an interesting effect: the Stark shift of the hyperfine states in these molecules depends on the molecular orientation. This effect can be used as the basis of an experimental scheme that could lead to a significantly improved measurement of the electron electric dipole moment.

We acknowledge support from NSERC, CFI, CRC, and a York Research Chair.